\documentstyle[titlepage,12pt]{article}
\begin{document}
\renewcommand{\theequation}{\thesection.\arabic{equation}}
\title{Semi-Classical Quantization of Circular Strings in de Sitter and
anti de Sitter Spacetimes}
\author{H.J. de Vega \thanks{Laboratoire de Physique
Th\'{e}orique et
Hautes Energies, Laboratoire Associ\'{e} \hspace*{6mm}au CNRS
UA280,  Universit\'{e} de Paris VI et VII, Tour 16,
1er \'{e}tage, 4, Place \hspace*{6mm}Jussieu,
75252 Paris cedex 05, France.} $^,$
\thanks{Isaac Newton Institute, Cambridge, CB3 0EH, United
Kingdom}, A.L. Larsen\thanks{Observatoire de Paris,
DEMIRM. Laboratoire Associ\'{e} au CNRS
UA 336, Ob- \hspace*{6mm}servatoire de Paris et
\'{E}cole Normale Sup\'{e}rieure. 61, Avenue
de l'Observatoire, \hspace*{6mm}75014 Paris, France.}
and
N. S\'{a}nchez $^3$}
\maketitle
\begin{abstract}
We compute the {\it exact} equation of state of circular strings in the
(2+1) dimensional de Sitter (dS) and anti de Sitter (AdS)
spacetimes, and analyze
its properties for the different (oscillating, contracting and expanding)
strings. The string equation of state has the perfect fluid form
$P=(\gamma-1)E,$ with the pressure and energy expressed closely and completely
in terms of elliptic functions, the instantaneous coefficient $\gamma$
depending on the elliptic modulus. We semi-classically quantize the oscillating
circular strings. The string mass is $m=\sqrt{C}/(\pi H\alpha'),\;C$
being the Casimir operator, $C=-L_{\mu\nu}L^{\mu\nu},$ of the $O(3,1)$-dS
[$O(2,2)$-AdS] group, and $H$ is the Hubble constant. We find
$\alpha'm^2_{\mbox{dS}}\approx 5.9n,\;(n\in N_0),$
and a {\it finite} number of states
$N_{\mbox{dS}}\approx 0.17/(H^2\alpha')$ in de Sitter spacetime;
$m^2_{\mbox{AdS}}\approx 4H^2n^2$ (large $n\in N_0$) and
$N_{\mbox{AdS}}=\infty$ in anti de Sitter spacetime. The level spacing
grows with $n$ in AdS spacetime, while is approximately constant (although
larger than in Minkowski spacetime) in dS spacetime. The massive states
in dS spacetime decay through tunnel effect and the semi-classical
decay probability is computed. The semi-classical
quantization of {\it exact} (circular) strings and the canonical
quantization of generic string
perturbations around the string center of mass strongly agree.
\end{abstract}

\section{Introduction and Results}

The systematic investigation of string dynamics  in curved spacetimes
started in Ref.\cite{plb}, has revealed new insights and
new physical phenomena with respect to string propagation in flat spacetime
(and with respect to quantum fields in curved spacetime) \cite{eri}.
These results  are relevant both for fundamental (quantum)
 strings and for cosmic strings, which behave, essentially,
in a classical  way.

Among the cosmological backgrounds, de Sitter  spacetime occupies a special
place. On one hand, it is relevant for inflation, and on the
other hand, string propagation turns out to be particularly interesting there
\cite{plb}-\cite{dls}.

Recently, a novel feature for strings in de Sitter spacetime
was found: exact multi-string solutions. That is, one single
world-sheet generically describes two  strings \cite{sash}, several
strings \cite{cdms},
  and even {\it infinitely many}  \cite{dls} (different and
independent) strings.

Circular strings are specially suited for detailed
investigation. Since the string equations of motion become separable, one has
to deal with  non-linear ordinary differential equations instead of
 non-linear  partial differential equations. In order to obtain
generic non-circular
string solutions the full power of the inverse scattering method is
needed in de Sitter spacetime \cite{cdms}.

Cosmological spacetimes are not Ricci flat and hence they are not
string vacua even at first order in $\alpha'$. Strings are there
non-critical and quantization will presumably lead to features like
ghost states. No definite answer is available by now to such
conformal anomaly effects.

We think it is important in this context to investigate the quantum
aspects in the semi-classical regime,
where anomaly effects are practically irrelevant.
Semi-clasically, in this context, means the regime in which $H^2\alpha'<<1,$
where $H$ is the Hubble constant.
We proceed in this paper to semi-classically quantize time-periodic string
solutions in de Sitter and anti de Sitter spacetimes after dealing
with Minkowski spacetime as an instructive exercise. Time-periodic string
solutions here include all the circular string solutions in Minkowski and anti
de Sitter spacetimes, as well as the oscillating string solutions in de Sitter
spacetime.

In this paper, we also complete the physical characterization of all circular
string solutions found recently in de Sitter \cite{dls} and in
anti de Sitter spacetimes \cite{ads}, by computing
the corresponding equations of state from the exact
string dynamics.

The circular string solutions in  de Sitter and anti de Sitter
spacetimes depend on an elliptic modulus $k$ and $\bar{k}$,
respectively. From the exact solutions we find their energy momentum
tensor. It turns out to have the perfect fluid form with an equation of state
\begin{equation}
P=(\gamma-1)E,
\end{equation}
where $\gamma$ in general is time-dependent and depends on the elliptic
modulus as well. We analyze the equation of state for all circular string
solutions in de Sitter and anti de Sitter spacetimes. In de Sitter
spacetime, for strings expanding from zero radius towards infinity, the
equation of state changes continuosly from the ultra-relativistic matter-type
when $r\approx 0,\;\;P=+E/2,$ (in $2+1$ dimensions) to the unstable
string-type, $P=-E/2,$ when $r\rightarrow\infty.$ On the other hand, for an
oscillating stable string in de Sitter spacetime, $\gamma$
oscillates between $\gamma(r=0)=3/2$ and
$\gamma(r=r_{\mbox{max}})=1/2+k^2/(1+k^2),$ where $k\in[0,\;1].$
Averaging over one oscillation period, the pressure {\it vanishes}.
That is, these stable string solutions actually describe {\it cold matter}.

In anti de Sitter spacetime, only
oscillating (stable) circular string solutions exist.
We find that $\gamma$ oscillates between $\gamma(r=0)=3/2$ and
$\gamma(r=r_{\mbox{max}})=1/2,$ i.e. the equation of state "oscillates"
between $P=+E/2$ and $P=-E/2.$ This is similar to the situation in flat
Minkowski spacetime. When averaging over an oscillation period in anti
de Sitter spacetime, we find that $\gamma$
takes values from $1$ to $1+1/\pi^2$ for the allowed range of the elliptic
modulus. That is, the average pressure over one oscillation period is
always {\it positive} in anti de Sitter spacetime.

In general, positive pressure characterizes the regime in which the string
radius is small relative to the string maximal size, while negative
pressure is characteristic for the regime in which the string radius is large.
In Minkowski spacetime, the two regimes are of equal "size", in the sense that
the average pressure is identically zero. The influence of the spacetime
curvature is among other effects, to modify the relative "size" of these two
regimes.

In order to semi-classically quantize these string solutions, we
compute the classical action $S_{\mbox{cl}}$ as a function of the
string mass $m$:
\begin{equation}
m\equiv -\frac{dS_{\mbox{cl}}}{dT}
\label{extrem}
\end{equation}
where we choose $T$ as the period in the {\bf physical} time variable
(in general different from the world-sheet time).
The quantization condition takes the form:
\begin{equation}
W(m)\equiv S_{\mbox{cl}}(T(m))+m\;T(m)\;=\;2\pi n,\quad n \in N_{0}
\end{equation}
In Minkowski spacetime, this formula reproduces the exact mass
spectrum except for  the intercept [see Eq.(\ref{sces})].

We find for de Sitter (anti de Sitter) spacetime that the mass is
exactly proportional to  the square-root of the Casimir operator
$C=-L_{\mu\nu}L^{\mu\nu}$ of the $O(3,1)$-de Sitter
[$O(2,2)$-anti de Sitter] group:
\begin{equation}
m = {\sqrt{C} \over {\pi  H \alpha'}}. \nonumber
\end{equation}
In Figures 4 and 5 we give parametric plots of
$H^2\alpha'W$ as a function of $H^2m^2\alpha'^2$ for $k\in[0,\;1]$
for  de Sitter spacetime and for
${\bar k}\in[0,\;1/\sqrt{2}\;[\;$  for anti
de Sitter spacetime, respectively.
A linear approximation turns out
to be rather accurate for de Sitter  spacetime:
\begin{equation}
\alpha'm^2_{\mbox{dS}}\approx 5.9\;n,\quad n\in N_{0}
\end{equation}
This is different from the mass spectrum in Minkowski spacetime.
The level spacing is however still approximately
constant, but the levels are more separated than in Minkowski spacetime.
In de Sitter spacetime
there is only a {\it finite} number of levels as can be seen from
Fig.4.  The number of
quantized circular string states can be estimated to be:
\begin{equation}\label{numm}
N_q \approx\frac{0.17}{H^2\alpha'}.
\end{equation}
It is interesting to compare this result with the number of particle
states obtained using canonical quantization \cite{plb}. One finds in
this way a  maximum number of states:
\begin{equation}
N_{\mbox{max}}\approx \frac{0.15}{H^2\alpha'},
\end{equation}
which is very close to the semi-classical value (\ref{numm}). It must be
noticed that in de Sitter spacetime, these states can {\it decay} quantum
mechanically due to the possibility of quantum mechanical tunneling through
the potential barrier, see Fig.1b. Semi-classically, the decay probability
is however highly suppressed for $H^2\alpha'<<1$ and for any value of the
elliptic modulus $k,$ except near $k=1$ where the barrier disappears,
and for which the tunneling probability is close to one.

In anti de Sitter spacetime arbitrary high mass states exist.
The quantization of the high mass states yields:
\begin{equation}
\alpha' m^2_{\mbox{AdS}}\approx 4H^2\alpha'\;n^2.
\end{equation}
Thus the (mass)$^2$ grows like $n^2$ and the level spacing grows
proportionally to $n.$ This is a completely different behaviour as compared
to Minkowski spacetime where the level spacing is constant.
Exactly the same result was
recently found, using canonical quantization of generic strings
in anti de Sitter spacetime \cite{ads,all}. The
physical consequences, especially
the non-existence of a critical string temperature (Hagedorn
temperature), of this kind of behaviour
is discussed in detail in Ref.\cite{all}.

For both de Sitter and anti de Sitter spacetimes we find thus a very
strong agreement between the results obtained using canonical quantization,
based on generic string solutions (string perturbation approach), and the
results obtained using the semi-classical approach, based on oscillating
circular string configurations.

This paper is organized as follows: In Section 2 we
describe the time-periodic string solutions in Minkowski, de
Sitter and anti de Sitter spacetimes. We derive the corresponding equations of
state and give the physical interpretation in the various regimes for the
different kinds of string solutions (oscillating and non-oscillating).
In Section 3 we proceed to quantize the oscillating strings
semi-classically, deriving the
quantum mass spectrum, and we compare with the results obtained using
canonical quantization. A summary of our results and conclusions is
presented in Section 4 and in Tables I, II.
\section{Periodic String Solutions and their  Physical Interpretation}
\setcounter{equation}{0}
The evolution of circular strings in curved spacetimes has recently been
discussed from both gravitational and cosmological points of view [3-9].
For completeness and
comparison we first consider flat Minkowski spacetime. We then
investigate the string dynamics in de Sitter spacetime (negative local gravity)
and finally consider anti de Sitter spacetime (positive local gravity).
We investigate  the effects of positive and negative local gravity
in the energy-momentum tensor of such circular strings.

The string equations of motion and constraints for a circular string are
most easily solved using static coordinates:
\begin{equation}
ds^2=-a(r)dt^2+\frac{dr^2}{a(r)}+r^2 d\phi^2,
\end{equation}
For simplicity we consider the string dynamics in a 2+1 dimensional spacetime.
All our solutions can however be embedded in a higher dimensional spacetime,
where they will describe plane circular strings. The circular string ansatz
($t=t(\tau),\;\;r=r(\tau),\;\;\phi=\sigma$) leads, after one integration,
to the following set of first order ordinary differential equations:
\begin{equation}
\dot{t}=\frac{\sqrt{b}\,\alpha'}{a(r)},
\label{ciruno}
\end{equation}
\begin{equation}
\dot{r}^2+r^2 a(r)=b\,\alpha'^2,
\label{cirdos}
\end{equation}
where $b$ is a non-negative integration constant with the dimension of
(mass)$^2.$ The left hand side of Eq.(2.3) is in the form of ``kinetic'' $+$
``potential'' energy. The potential is given by $V(r)=r^2\, a(r)$ ,
where:
\begin{eqnarray}
a(r) &=& 1 \quad{\rm for~Minkowski~ spacetime},\nonumber \\
a(r)&=&1-H^2r^2 \quad{\rm for~de~ Sitter~ spacetime},\nonumber \\
a(r)&=&1+H^2r^2 \quad{\rm for~anti~ de~ Sitter ~spacetime}.\nonumber
\end{eqnarray}
Properties like
energy and pressure of the strings are more conveniently discussed in
comoving (cosmological) coordinates:
\begin{equation}
ds^2=-(dX^0)^2+a^2(X^0)\;\frac{dR^2+R^2 \,d\phi^2}{(1+\frac{k}{4}R^2)^2},
\end{equation}
including as special cases Minkowski, de
Sitter and anti de Sitter
spacetimes:
\begin{eqnarray}
a(X^0) &=& 1,\;\;k=0 \quad{\rm for~Minkowski~ spacetime},\nonumber \\
a(X^0)&=&e^{HX^0},\;\;k=0 \quad{\rm for~de~ Sitter~ spacetime},\nonumber \\
a(X^0)&=&\cos HX^0,\;\;k=-H^2
\quad{\rm for~anti~ de~ Sitter ~spacetime}.\nonumber
\end{eqnarray}
The spacetime energy-momentum tensor is given in $2+1$ dimensional
spacetime by:
\begin{equation}
\sqrt{-g}\,T^{\mu\nu}=\frac{1}{2\pi\alpha'}\int d\tau d\sigma\;(\dot{X}^\mu
\dot{X}^\nu-X'^\mu X'^\nu)\;\delta^{(3)}(X-X(\tau,\sigma)).
\end{equation}
After integration over a spatial volume
that completely encloses the string \cite{cos}, the
energy-momentum tensor for a circular string takes the form of a fluid:
\begin{equation}
T^\mu_\nu=\mbox{diag.}(-E,\;P,\;P),
\end{equation}
where, in the comoving coordinates (2.4):
\begin{equation}
E(X)=\frac{1}{\alpha'}\,\dot{X}^0,
\end{equation}
\begin{equation}
P(X)=\frac{1}{2\alpha'}\,\frac{a^2(X^0)}{(1+\frac{k}{4}R^2)^2}\;
\frac{\dot{R}^2-R^2}{\dot{X}^0},
\end{equation}
represent the string energy and pressure, respectively.
\subsection{Minkowski Spacetime}
In this case Eqs.(\ref{ciruno}), (\ref{cirdos}) are solved by:
\begin{equation}
r(\tau)=\sqrt{b}\,\alpha'\mid\cos\tau\mid,\;\;\;\;\;\;t(\tau)=
\sqrt{b}\,\alpha'\tau,
\label{cirmin}
\end{equation}
i.e. the string radius follows a pure harmonic motion with
period in world-sheet time $T_\tau=\pi.$
The energy and pressure, Eqs.(2.7), (2.8), are:
\begin{equation}
E=\sqrt{b},
\end{equation}
\begin{equation}
P=\frac{b\alpha'^2-2r^2}{2\sqrt{b}\alpha'}=
-\frac{\sqrt{b}}{2}\, \cos 2\tau.
\end{equation}
The energy is constant while the pressure depends on the string radius. For
$r\rightarrow 0\;\;(\tau\rightarrow \pi/2)$, that is when the string is
collapsed, we find the equation of state $P=E/2$ correspondig to
ultra-relativistic matter in $2+1$ dimensions. For
$r\rightarrow \sqrt{b}\,\alpha'\;
(\tau\rightarrow\pi)$, that is when the string has its maximal size, the
pressure is negative and we find $P=-E/2.$ This is the same equation of state
that was found for extremely unstable strings in inflationary universes
\cite{ven}. The circular string thus oscillates between these two limiting
types of equation of state. This illustrates that instantaneous
negative pressure
is a generic feature of strings, not only for  unstable strings in
inflationary universes, but even for stable oscillating strings in flat
Minkowski spacetime. Even in flat Minkowski spacetime we see that there is a
positive pressure regime (when the string radius is small relative to
its maximal size) and a negative pressure regime (when the string radius
is large). For the circular strings in Minkowski spacetime the two regimes are
of equal size in the sense that the average pressure equals zero (as can be
easily
shown by integrating Eq.(2.11) over a full period). The strings thus, in
average, obey an equation of state of the cold matter type. The influence of
the
curvature of spacetime is, among other effects, to change the relative
`size' of the positive pressure regime to the negative pressure regime, as
we will see in the following subsections.
\subsection{de Sitter Spacetime}
In de Sitter spacetime
the solution of Eqs.(2.2), (2.3) involves elliptic functions.
As can be seen from the potential, Fig.1b, the dynamics in de Sitter
spacetime is
completely different from the dynamics in Minkowski
and anti de Sitter spacetime.
The inflation of the background here gives rise to a finite barrier implying
the existence of oscillating strings as well as
contracting and expanding strings, whose
exact dynamics was discussed in detail in Ref.\cite{dls}. The energy and
pressure have been discussed in Ref.\cite{dls}. In this subsection we further
analyze the string energy and pressure in de Sitter spacetime.
The coordinate transformation relating the line elements (2.1) and (2.4), in
the case of de Sitter spacetime, is
given by:
\begin{equation}
X^0=\frac{1}{2H}\,\log\mid 1-H^2r^2\mid + \, t,
\end{equation}
\begin{equation}
R=\frac{re^{-Ht}}{\sqrt{1-H^2r^2}}.
\end{equation}
The energy and pressure then take the form:
\begin{equation}
H\alpha'E=\frac{H^2r\dot{r}-H\sqrt{b}\alpha'}{H^2r^2-1},
\end{equation}
\begin{equation}
H\alpha'P=\frac{(H^4r^4-2H^2r^2-H^2b\alpha'^2)H^2r\dot{r}+
(3H^4r^4-2H^2r^2+H^2b\alpha'^2)H\sqrt{b}\alpha'}{2(1-H^2r^2)(H^4r^4+
H^2b\alpha'^2)}.
\end{equation}
Both the energy and the pressure now depend on the string radius $r$ and the
velocity $\dot{r}.$ The latter can however be eliminated using Eq.(2.3).

Let us first consider a string expanding from $r=0$ towards infinity.
This corresponds to a string with $\dot{r}>0$ and $H^2b\alpha'^2>1/4$, see
Fig.1b. For $r=0$ we find $E=\sqrt{b}$ and $P=\sqrt{b}/2,$ thus the equation
of state $P=E/2.$ This is the same result as in Minkowski spacetime, i.e. like
ultra-relativistic matter. As the string expands, the energy soon starts to
increase while the pressure starts to decrease and becomes negative, see
Fig.2. For $r\rightarrow\infty$ we find $E=r/\alpha'$ and $P=-r/(2\alpha'),$
thus, not surprisingly, we have recovered the equation of state of extremely
unstable strings \cite{ven}, $P=-E/2.$

Now consider an oscillating string, i.e. a string
with $H^2b\alpha'^2\leq 1/4$ in the region to the left of the potential
barrier, Fig.1b. The equation of state near $r=0$ is the same as for the
expanding string, but now the string has a maximal radius:
\begin{equation}
Hr_{\mbox{max}}\equiv \nu=\sqrt{\frac{1-\sqrt{1-4H^2b\alpha'^2}}{2}}.
\end{equation}
For $r=r_{\mbox{max}}$ we find:
\begin{equation}
H\alpha'E=\frac{\nu}{\sqrt{1-\nu^2}}\equiv k,\;\;\;\;\;\;
H\alpha'P=\frac{k}{2}(-1+\frac{2k^2}{1+k^2}),
\end{equation}
corresponding to a perfect fluid type equation of state:
\begin{equation}
P=(\gamma-1)E,\;\;\;\;\;\;\;\;\gamma=\frac{k^2}{1+k^2}+\frac{1}{2}.
\end{equation}
Notice that $k\in[0,\;1]$ with $k=0$ decribing a string at the bottom of the
potential while $k=1$ decribes a string oscillating between $r=0$ and the
top of the potential barrier (in this case the string actually only makes
one oscillation \cite{sash,dls}). For
$k<<1$ the equation of state (2.18) near the
maximal radius reduces to $P=-E/2$ which is the same result we found in
Minkowski spacetime. In the other limit $k\rightarrow 1$ we find, however,
$P=0$ corresponding to cold matter.

For the oscillating
strings we can also calculate the average values of energy and pressure by
integrating Eqs.(2.14), (2.15) over a full period, Fig.3. Since
a $\tau$-integral can be converted into
a $r$-integral, using Eq.(2.3), the average values can be obtained without
using the exact $\tau$-dependence of the string radius [the solution of
Eq.(2.3)]. The average energy becomes:
\begin{equation}
H\alpha'<E>=\frac{2k}{T_{\tau}\sqrt{1+k^2}}\,\Pi(\frac{k^2}{1+k^2},\;k),
\end{equation}
where $T_{\tau}$ is the period in the world-sheet time $\tau$
and $\Pi$ is the complete elliptic integral of the
third kind. The average pressure can be expressed here as a
combination of complete elliptic integrals of the third
kind. Numerical evaluation gives zero with high accuracy. We have
checked in addition  that the first orders in the expansion in the
elliptic modulus $k$ identically vanish.
We conclude that the average pressure
is zero as in Minkowski spacetime. Thus, in average the
{\it oscillating strings}
describe {\it cold matter}.
\subsection{Anti de Sitter Spacetime}
As an example of a FRW-universe with positive local gravity we now consider
anti de Sitter spacetime. Here the circular string potential goes to infinity
even faster than in Minkowski spacetime, see Fig.1c, so we can only have
oscillating string solutions. The maximal string radius is given by:
\begin{equation}
Hr_{\mbox{max}}\equiv\bar{\nu}=\sqrt{\frac{-1+\sqrt{1+4H^2b\alpha'^2}}{2}},
\end{equation}
that can take any non-negative value. The comoving coordinates, Eq.(2.4),
are in the case of anti de Sitter spacetime given by:
\begin{equation}
HX^0=\pm\arccos\sqrt{(1+H^2r^2)\cos^2Ht-H^2r^2},
\end{equation}
\begin{equation}
HR=\frac{2}{Hr}[\sqrt{1+H^2r^2}\cos Ht-\sqrt{(1+H^2r^2)\cos^2Ht-H^2r^2}\;].
\end{equation}
The energy and pressure can be written as:
\begin{equation}
H\alpha' E=\frac{H^2\,r\dot{r}\sin Ht+H\sqrt{b}\alpha'
\cos Ht}{\sqrt{1+H^2r^2}
\sqrt{(1+H^2r^2)\cos^2Ht-H^2r^2}},
\end{equation}
\begin{equation}
H\alpha' P=\frac{[H\dot{r}\cos Ht+H^2\sqrt{b}\alpha' r\sin Ht]^2-
H^2r^2(1+H^2r^2)[(1+H^2r^2)\cos^2Ht-H^2r^2]}
{2[H^2r\dot{r}\sin Ht+H\sqrt{b}\alpha'
\cos Ht]\sqrt{1+H^2r^2}
\sqrt{(1+H^2r^2)\cos^2Ht-H^2r^2}}.
\end{equation}
Using Eqs.(2.2), (2.3) we can then write down the energy and pressure
explicitly as functions of the string radius $r,$ only.
For the present purposes it is however sufficient to consider some
limiting cases. It is convenient to introduce the parameter $\bar{k}$:
\begin{equation}
\bar{k}\equiv\frac{\bar{\nu}}{\sqrt{1+2\bar{\nu}^2}}\in[0,\;\sqrt{1/2}\;[
\end{equation}
For $\bar{k}\rightarrow 0$ the string oscillates near the bottom of the
potential, while the other extreme corresponds to $\bar{k}\rightarrow
\sqrt{1/2}.$ For $r=0$ we find:
\begin{equation}
H\alpha'E=\frac{\bar{k}\sqrt{1-\bar{k}^2}}{1-2\bar{k}^2}=2H\alpha'P,
\end{equation}
thus the ultra-relativistic matter equation of state.

For $r=r_{\mbox{max}}$
we find:
\begin{equation}
H\alpha'E=\frac{\bar{k}}{\sqrt{1-2\bar{k}^2}},\;\;\;\;\;\;
H\alpha'P=-\frac{\bar{k}}{2\sqrt{1-2\bar{k}^2}},
\end{equation}
corresponding to the equation of state $P=-E/2.$ This is exactly as in
Minkowski spacetime: the string oscillates between the two limiting types
of equation of state, $P=E/2$ and $P=-E/2.$ A new
phenomenon appears, however, when we calculate the average values over a full
period. The average energy becomes:
\begin{equation}
H\alpha'<E>=\frac{2\bar{k}}{T_\tau}\sqrt{\frac{1-2\bar{k}^2}{1-\bar{k}^2}}\,\Pi
(\frac{\bar{k}^2}{1-\bar{k}^2},\;\bar{k}),
\end{equation}
where $T_\tau$ is the period in the world-sheet time $\tau$
and $\Pi$ is the complete elliptic integral of the
third kind. The
average pressure for oscillating strings in anti de Sitter
spacetime is, contrary to Minkowski and de Sitter spacetime, non-zero.
No simple analytic expression for it has been
found for arbitrary $\bar{k}.$ The equation of state is of perfect
fluid type $<P>=(\gamma-1)<E>$ where $\gamma$ depends on $\bar{k}.$
Approximate results can
be obtained in the two extreme limits. For $\bar{k}<<1$ we find:
\begin{equation}
H\alpha'<P>=\frac{\bar{k}^3}{32}+{\cal O}(\bar{k}^5),
\end{equation}
while Eq.(2.28) gives:
\begin{equation}
H\alpha'<E>=\bar{k}+{\cal O}(\bar{k}^3),
\end{equation}
and therefore:
\begin{equation}
\gamma=1+\frac{\bar{k}^2}{32}+{\cal O}(\bar{k}^4).
\end{equation}
In the limit $\bar{k}\rightarrow\sqrt{1/2}$ we find:
\begin{equation}
H\alpha'<P>=\frac{1}{2\pi K(\sqrt{1/2}\;)\sqrt{1-2\bar{k}^2}}+
{\cal O}(\sqrt{1-2\bar{k}^2}\;),
\end{equation}
and from Eq.(2.28):
\begin{equation}
H\alpha'<E>=\frac{\pi}{2K(\sqrt{1/2}\;)\sqrt{1-2\bar{k}^2}}+
{\cal O}(\sqrt{1-2\bar{k}^2}\;),
\end{equation}
where $K$ is the complete elliptic integral of the first kind.
This corresponds to:
\begin{equation}
\gamma=1+1/\pi^2+{\cal O}(1-2\bar{k}^2).
\end{equation}
Numerical evaluation of  $<E>$ and $<P>$ shows that $\gamma$
monotonically grows from  $\gamma=1$ till  $\gamma=1+1/\pi^2$ when
$\bar{k}$ grows from zero to $1/\sqrt2,$ so that the average pressure
is always positive.

This concludes our analysis of the various types of equation
of state for circular strings in Minkowski, de Sitter and anti de Sitter
spacetimes. The results are summarized in Table I.
\section{Semi-Classical Quantization}
\setcounter{equation}{0}
In this section we perform a semi-classical quantization of the circular
string configurations discussed in the previous section. We use an approach
developed in field theory by Dashen et. al. \cite{dhn,hm}, based
on the stationary phase approximation of
the partition function. The method can be only used for time-periodic
solutions of
the classical equations of motion. In our string problem, these
solutions however, include all the circular
string solutions in Minkowski and in anti de Sitter spacetimes, as well as the
oscillating circular strings ($H^2b\alpha'^2\leq 1/4,$ c.f. Subsection 2.2)
in de Sitter spacetime.

The result of the stationary phase integration is expressed in terms of
the function:
\begin{equation}
W(m)\equiv S_{\mbox{cl}}(T(m))+m\;T(m),
\end{equation}
where $S_{\mbox{cl}}$ is the action of the classical solution, $m$ is the
mass and
the period $T(m)$ is implicitly given by:
\begin{equation}
\frac{dS_{\mbox{cl}}}{dT}=-m.
\label{extre}
\end{equation}
In string theory we must choose $T$ to be the period in a physical
time variable. For example, when a light cone gauge exists, $T$ is the
period in $X^0=\alpha' p\tau.$
The bound state quantization condition then becomes \cite{dhn,hm}:
\begin{equation}
W(m)=2\pi n,\quad n \in N_{0}
\label{concua}
\end{equation}
for $n$ `large'. The method has been successfully used in many cases from
quantum mechanics to quantum field theory. For integrable
field theories the semi-classical quantization happens in fact,
to be exact. It must be noticed that string
theory in de Sitter spacetime is exactly integrable \cite{san}.

To demonstrate the method and to fix the normalization we first consider
the circular strings in flat Minkowski spacetime. We then perform the same
analysis for de Sitter and anti de Sitter spacetimes, and then after,
we compare with
approximate results obtained using canonical quantization \cite{plb,all}.
\subsection{Minkowski Spacetime}
The string action in the conformal gauge in Minkowski spacetime is given by:
\begin{equation}
S=\frac{1}{2\pi\alpha'}\int_0^{T}d\sigma\int_0^{T} d\tau\;g_{\mu\nu}\;
(\dot{X}^\mu\dot{X}^\nu-X'^\mu X'^\nu).
\label{acT}
\end{equation}
where the world-sheet coordinate $\sigma$ runs from $0$ to $T$. That is,
$$
X^{\mu}(\sigma+T,\tau) = X^{\mu}(\sigma,\tau)
$$
In this parametrization the circular string solution
(\ref{cirmin}) takes the form
\begin{eqnarray}
X^0 &=& A\,\tau \\
X^1 &=& {A\,T \over {2\pi}} \, \cos\left({{2\pi\sigma}\over T}\right)
\, \cos\left({{2\pi\tau}\over T}\right) \\
X^2 &=& {A\,T \over {2\pi}} \, \sin\left({{2\pi\sigma}\over T}\right)
\, \cos\left({{2\pi\tau}\over T}\right)
\label{repc}
\end{eqnarray}
where $A$ is an arbitrary constant [$A=\sqrt{b}\alpha'$ in the notation
of Eq.(2.9)] and $T$ became the period in the
$\tau$ variable too.
For this solution in Minkowski spacetime, we find from Eq.(\ref{acT}):
\begin{equation}
S_{\mbox{cl}}=-\frac{A^2}{2\pi\alpha'}\,T^2
\end{equation}
Equation (\ref{extre}) then takes the form:
\begin{equation}
M=\frac{A^2}{\pi\alpha'}\,T
\end{equation}
and  then the quantization condition (\ref{concua}) yields:
\begin{equation}
M^2=\frac{4 A^2}{\alpha'}\;n, \quad n\in N_{0}
\end{equation}
We must identify the mass with the  variable conjugated to
the physical time $X^0$. Since $M$ is conjugated to $\tau$ and $X^0
= A\,\tau$ , $m \equiv M/A $ is the string mass. Therefore the
semiclassical string spectrum results:
\begin{equation}
\alpha'\, m^2 = 4\,  n, \quad n\in N_{0}
\label{sces}
\end{equation}
If we subtract the intercept $-4$ in Eq.(\ref{sces})
this is the well-known (exact) mass formula
for closed bosonic strings in flat Minkowski spacetime.
\subsection{de Sitter Spacetime}
Using the notation introduced in Eqs.(2.16), (2.17), the oscillating strings
in de Sitter spacetime are given by \cite{dls}:
\begin{equation}
H r(\tau)=\frac{k}{\sqrt{1+k^2}}\mid\mbox{sn}
[\frac{\tau}{\sqrt{1+k^2}},\;k]\mid.
\end{equation}
Eq.(2.2) is then integrated to:
\begin{equation}
H t(\tau)=\frac{k}{\sqrt{1+k^2}}\Pi(\frac{k^2}{1+k^2},\;\frac{\tau}
{\sqrt{1+k^2}},\;k).
\end{equation}
The period in comoving time, which from Eq.(2.12) equals the period
in static coordinate time, is then given by:
\begin{equation}
H T=\frac{2k}{\sqrt{1+k^2}}\Pi(\frac{k^2}{1+k^2},\;k).
\end{equation}
Notice that the expressions for the periods in the physical time $X^0$
and in the
world-sheet time $\tau$ are different. The period in the physical time can be
further rewritten in terms of incomplete elliptic integrals of the first and
second kinds:
\begin{equation}
H T=\frac{2kK(k)}{\sqrt{1+k^2}}+2K(k)E(\phi,k)-2E(k)F(\phi,k),
\end{equation}
where:
\begin{equation}
\phi=\arcsin\frac{1}{\sqrt{1+k^2}}.
\end{equation}
The classical action over one period becomes:
\begin{equation}
S_{\mbox{cl}}=\frac{4}{H^2\alpha'}\;\frac{E(k)-K(k)}{\sqrt{1+k^2}},
\end{equation}
A straightforward calculation gives:
\begin{equation}
H\frac{dT}{dk}=-\frac{2}{\sqrt{1+k^2}}
\left[K(k)-\frac{2E(k)}{1-k^2}\right],
\end{equation}
as well as:
\begin{equation}
\frac{dS_{\mbox{cl}}}{dk}=\frac{4}{H^2\alpha'}\frac{k}{(1+k^2)^{3/2}}
\left[K(k)-\frac{2E(k)}{1-k^2}\right].
\end{equation}
Then, identifying the string mass $m$ as the conjugate
to the comoving time $X^0,$ Eq.(3.2) leads to:
\begin{equation}
m=\frac{2}{H\alpha'}\frac{k}{1+k^2}.
\end{equation}
The string solutions in de Sitter spacetime enjoy conserved quantities
associated with the $O(3,1)$ rotations on the hyperboloid. Using
hyperboloid coordinates, the only non-zero component for the circular
string solutions under consideration here,
is given by $L_{10}=-L_{01}=\sqrt{C}\;$  \cite{dls}:
\begin{equation}
L_{10}=\sqrt{C}=2\pi\frac{k}{1+k^2},\nonumber
\end{equation}
where $C=-L_{\mu\nu}L^{\mu\nu}$ is the Casimir operator of the group.
Hence, the mass is exactly {\bf linear} in $\sqrt{C}$
\begin{equation}
m = {\sqrt{C} \over {\pi  H \alpha'}}. \nonumber
\end{equation}
The physical meaning of such type of `linear' Regge trajectory
deserves further investigation.

The quantization condition (3.3) finally gives:
\begin{equation}
W=\frac{4}{H^2\alpha'\sqrt{1+k^2}}\left[E(k)-\frac{K(k)}{1+k^2}+
\frac{k[K(k)E(\phi,k)-E(k)F(\phi,k)]}{\sqrt{1+k^2}}\right]=2\pi n.
\end{equation}
This equation determines a quantization of the parameter $k,$ which by
Eq.(3.20) gives a quantization of the mass. A full parametric plot of
$H^2\alpha'W$ as a function of $H^2m^2\alpha'^2$ for
$k\in[0,\;1]$ is shown in Fig.4. In the
whole $k-$range a good approximation is provided by the line connecting
the two end-points:
\begin{eqnarray}
W\hspace*{-2mm}&=&\hspace*{-2mm}[2\sqrt{2}-2\log (1+\sqrt{2})]
m^2\alpha' \nonumber\\
\hspace*{-2mm}&\approx &\hspace*{-2mm}1.06\;m^2\alpha',
\end{eqnarray}
and the quantization of the mass becomes:
\begin{equation}
\alpha'm^2\approx 5.9\;n,\quad n\in N_{0}
\end{equation}
This is different from the result obtained in Minkowski spacetime.
The level spacing is however still approximately
constant, but the levels are more separated than in Minkowski spacetime.
In de Sitter spacetime
there is only a {\it finite} number of levels as can be seen from
Fig.4. This is due to the {\it finite}
height of the potential barrier. The number of
quantized circular string states is easily estimated using
Eqs.(3.23), (3.24) and
Fig.4:
\begin{equation}
N_q \approx 1+\mbox{Int}\left( \frac{2\sqrt{2}-2\log (1+\sqrt{2})}
{2\pi H^2\alpha'}\right).
\end{equation}
For
$H^2\alpha'<<1,$ which is clearly fulfilled in most interesting cases, we
get:
\begin{equation}
N_q \approx\frac{0.17}{H^2\alpha'}.
\end{equation}
It should be stressed, however, that these states are not truely stable
stationary states because of the possibility of quantum mechanical tunneling
through the barrier. The probability of decay is given by:
\begin{equation}
{\cal T}\propto e^{-S_E},
\end{equation}
where $S_E$ is the Euclidean action of the classical solution in the
classically forbidden region. Defining $t=it_E,\;\tau=i\tau_E$ and
$S_E=iS,$ we find from Eqs.(2.2)-(2.3):
\begin{equation}
\frac{dt_E}{d\tau_E}=\frac{\sqrt{b}\;\alpha'}{1-H^2r^2},
\end{equation}
\begin{equation}
\left(\frac{dr}{d\tau_E}\right)^2=r^2(1-H^2r^2)-b\alpha'^2.
\end{equation}
Then the Euclidean action takes the form:
\begin{equation}
S_E=\frac{1}{2\pi\alpha'}\int_0^{2\pi}d\sigma\int_{period}
\hspace*{-5mm}d\tau_E
\left\{ (1-H^2r^2)\left(\frac{dt_E}{d\tau_E}\right)^2-\frac{1}{1-H^2r^2}
\left(\frac{dr}{d\tau_E}\right)^2
-r^2\left(\frac{d\phi}{d\sigma}\right)^2\right\}.
\end{equation}
The integral can be expressed in terms of complete elliptic integrals of
second and third kinds:
\begin{equation}
S_E=\frac{4}{H^2\alpha'}\;\frac{E(k')-\Pi\left(-(k'/k)^2,\;k'\right)}{\sqrt{1+k^2}},
\end{equation}
where $k'=\sqrt{1-k^2}.$ For $H^2\alpha'<<1$ the quantum
mechanical tunneling is highly suppressed for any value of $k$ except
near $k=1,$ as
follows by analyzing Eq.(3.32) in a little more detail.
For $k\rightarrow 1,$ where the barrier disappears,
the decay probability becomes unity. For $k\rightarrow 0,$ near the
bottom of the potential, the decay probability is:
\begin{equation}
{\cal T}_{k\rightarrow 0}\propto \mbox{exp}\left[-\frac{4}{H^2\alpha'}\left\{
1-\frac{\pi}{2}k-\frac{k^2}{2}\log\frac{4}{k}+{\cal O}(k^2)\right\}\right].
\end{equation}
For $k$ identically zero, the decay process can be interpreted
as a creation of strings with probability
${\cal T}_{k=0}\propto \mbox{exp}[-\frac{4}{H^2\alpha'}].$  This $k=0$-term
coincides with the result found by Basu, Guth and Vilenkin \cite{bgv} in
the context of a cosmic string nucleation scenario.

Let us now return to the number of states, Eq.(3.27). It is interesting to
compare the results here with the results obtained using canonical quantization
of generic strings. By using a string
perturbation series approach for $H^2\alpha'<<1,$
it was shown by de Vega and S\'{a}nchez
\cite{plb}
 that the mass formula in de Sitter spacetime takes the form:
\begin{equation}
\alpha'm^2=24\sum_{n>0}\frac{2n^2-H^2m^2\alpha'^2}
{\sqrt{n^2-H^2m^2\alpha'^2}}+\sum_{n>0}\frac{2n^2-H^2m^2\alpha'^2}
{\sqrt{n^2-H^2m^2\alpha'^2}}\sum_R [(\alpha^R_n)^{\dag}
\alpha^R_n+(\tilde{\alpha}^R_n)^{\dag}\tilde{\alpha}^R_n],
\end{equation}
where:
\begin{equation}
[\alpha^R_n,\;(\alpha^R_n)^{\dag}]=[\tilde{\alpha}^R_n,\;(\tilde{\alpha}^R_n)^
{\dag}]=1.
\end{equation}
It follows that real mass solutions can only be defined up to some maximal
mass of the order $\alpha'm^2\approx 1/(H^2\alpha').$ To be a little more
specific consider physical states in the form:
\begin{equation}
(\tilde{\alpha}^{R_1}_1)^{\dag}(\alpha^{S_1}_1)^{\dag},.\;.\;.\;.\;.\;.,
(\tilde{\alpha}^{R_N}_1)^{\dag}(\alpha^{S_N}_1)^{\dag}\mid 0>,
\end{equation}
with mass implicitly given by:
\begin{equation}
\alpha'm^2=24\sum_{n>0}\frac{2n^2-H^2m^2\alpha'^2}
{\sqrt{n^2-H^2m^2\alpha'^2}}+2N\frac{2-H^2m^2\alpha'^2}{\sqrt
{1-H^2m^2\alpha'^2}}.
\end{equation}
For $H^2\alpha'<<1$ we find real mass solutions to this equation only
for:
\begin{equation}
N\leq N_{\mbox{max}}\approx \frac{0.15}{H^2\alpha'}.
\end{equation}
Thus, along a trajectory in the Regge plot, we find only $N_{\mbox{max}}$
states. This is the relevant quantity to be compared with the number of
{\it exact} circular string states $N_q$ in the potential, and the two
numbers are in fact
of the same order, compare with Eq.(3.27).
\subsection{Anti de Sitter Spacetime}
The calculations here are very similar to the calculations of Subsection 3.2,
but the results will turn out to be completely different.
In the notation of Eqs.(2.20), (2.25) the oscillating strings
in anti de Sitter spacetime are given by \cite{ads}:
\begin{equation}
Hr(\tau)=\frac{\bar{k}}{\sqrt{1-2\bar{k}^2}}\mid\mbox{cn}
[\frac{\tau}{\sqrt{1-2\bar{k}^2}},\;\bar{k}]\mid.
\end{equation}
In this case Eq.(2.2) leads to:
\begin{equation}
Ht(\tau)=\bar{k}\sqrt{\frac{1-2\bar{k}^2}{1-\bar{k}^2}}
\;\Pi\left(\frac{\bar{k}^2}{1-\bar{k}^2},
\;\frac{\tau}{\sqrt{1-2\bar{k}^2}},\;\bar{k}\right).
\end{equation}
Also in anti de Sitter spacetime, from Eq.(2.21), the period in comoving
time equals the period in static coordinate time:
\begin{equation}
HT=2\bar{k}\sqrt{\frac{1-2\bar{k}^2}{1-\bar{k}^2}}
\;\Pi(\frac{\bar{k}^2}{1-\bar{k}^2},\;\bar{k}).
\end{equation}
It is rewritten in terms of incomplete elliptic integrals of the first and
second kinds:
\begin{equation}
HT=\pi+2\bar{k}K(\bar{k})\sqrt{\frac{1-2\bar{k}^2}{1-\bar{k}^2}}
-2K(\bar{k})E(\bar{\phi},\bar{k'})+
2[K(\bar{k})-E(\bar{k})]F(\bar{\phi},\bar{k'}),
\end{equation}
where:
\begin{equation}
\bar{\phi}=\arcsin\frac{\sqrt{1-2\bar{k}^2}}{1-\bar{k}^2},\;\;\;\;\;\;\;\;
\bar{k'}=\sqrt{1-\bar{k}^2}.
\end{equation}
The classical action over one period becomes:
\begin{equation}
S_{\mbox{cl}}=\frac{4}{H^2\alpha'}\;\frac{(1-\bar{k}^2)K(\bar{k})-E(\bar{k})}
{\sqrt{1-2\bar{k}^2}}.
\end{equation}
A straightforward calculation gives:
\begin{equation}
H\frac{dT}{d\bar{k}}=\frac{2}{\sqrt{(1-\bar{k}^2)(1-2\bar{k}^2)}}
[2E(\bar{k})-K(\bar{k})],
\end{equation}
as well as:
\begin{equation}
\frac{dS_{\mbox{cl}}}{d\bar{k}}=-\frac{4\bar{k}}{H^2\alpha'}
\;\frac{2E(\bar{k})-K(\bar{k})}{(1-2\bar{k}^2)^{3/2}}.
\end{equation}
The mass is obtained from Eq.(3.2):
\begin{equation}
m=\frac{2\bar{k}}{H\alpha'}\frac{\sqrt{1-\bar{k}^2}}{1-2\bar{k}^2}.
\end{equation}
As in de Sitter spacetime, we find here an
exact {\bf linear} relation between the mass and the square-root of the
Casimir operator $\sqrt{C}=L_{10}$ (in hyperboloid coordinates) of the group:
\begin{equation}
m = {\sqrt{C} \over {\pi  H \alpha'}}. \nonumber
\end{equation}
The quantization condition (3.3) finally gives:
\begin{eqnarray}
&W& \hspace*{-2mm}=
\frac{4}{H^2\alpha'}\left\{\frac{\pi}{2}\frac{\bar{k}\sqrt{1-\bar{k}^2}}
{1-2\bar{k}^2}+\frac{K(\bar{k})-E(\bar{k})}{\sqrt{1-2\bar{k}^2}}\right.
\nonumber\\
\hspace*{-2mm}&-&\hspace*{-2mm}\left.
\frac{\bar{k}\sqrt{1-\bar{k}^2}}{1-2\bar{k}^2}
\left[K(\bar{k})E(\bar{\phi},\bar{k'})-(K(\bar{k})-
E(\bar{k}))F(\bar{\phi},\bar{k'})\right]\right\}=2\pi n.~~~
\end{eqnarray}
This equation determines a quantization of the parameter $\bar{k},$ which by
Eq.(3.47) gives a quantization of the mass. A parametric plot of
$H^2\alpha'W$ as a function of $H^2m^2\alpha'^2$ for
$\bar{k}\in[0,\;1/\sqrt{2}[\;$
is shown in Fig.5. The curve continues
forever to the right (contrary to the case of de Sitter spacetime, Fig.4.),
so that arbitrarily
high mass states exist. In anti de Sitter spacetime, this is also
clear from the potential, Fig.1c. Asymptotically
($\bar{k}\rightarrow\sqrt{1/2}$) we find from Eqs.(3.47), (3.49):
\begin{equation}
m=\frac{1}{H\alpha'(1-2\bar{k}^2)}+{\cal O}(1),
\end{equation}
\begin{equation}
W=\frac{\pi}{H^2\alpha'(1-2\bar{k}^2)}+{\cal O}(\frac{1}{\sqrt{1-2\bar{k}^2}}),
\end{equation}
i.e.:
\begin{equation}
W\approx\frac{\pi}{H}m.
\end{equation}
The quantization of the high mass states then takes the form:
\begin{equation}
\alpha' m^2\approx 4H^2\alpha'\;n^2
\end{equation}
Thus the (mass)$^2$ grows like $n^2$ and the level spacing grows
proportionally to $n.$ This is a completely different behaviour as compared
to Minkowski spacetime where the level spacing is constant.
A similar result was
found recently, using canonical quantization of generic strings
in anti de Sitter spacetime \cite{all}.
The mass formula in anti de Sitter spacetime takes the form (3.34) but with
$H^2<0$ (reminiscent of the formal relation between de Sitter and anti de
Sitter line elements in static coordinates). Then, the square roots in
the denominators are well-defined for any value of $\alpha' m^2$ and arbitrary
high mass states can be constructed. By considering states of the form (3.36)
for very large $N$ ($N>>1/(H^2\alpha')$) it was shown that \cite{all}:
\begin{equation}
\alpha' m^2\approx 4H^2\alpha'\;N^2,
\end{equation}
in agreement with the result obtained here for circular strings,
Eq.(3.53). It should be noticed that the circular string oscillations in
anti de Sitter spacetime (and in de Sitter spacetime)
do not follow a pure harmonic
motion as in flat Minkowski spacetime. Since expressed in terms of Jacobi
elliptic functions they are in fact very precise superpositions of all
frequencies. The states Eq.(3.36), involving only one frequency, should
therefore not have exactly the same mass as a circular string, so we
can only expect a qualitative agreement for the results obtained using the
two different approaches, and that was indeed what we found.
\section{Conclusion}
We have computed {\it exactly} the equation of state of the circular string
solutions recently found in de Sitter \cite{dls} and anti de Sitter
\cite{ads} spacetimes. The string equation of state has the perfect fluid form
$P=(\gamma-1)E,$ with $P$ and $E$ expressed closely and completely in
terms of elliptic functions and the instantaneous parameter $\gamma$ depending
on the elliptic modulus. We have quantized the time-periodic (oscillating)
string solutions within the semi-classical (stationary phase approximation)
approach.

The main results of this paper are summarized in Tables I and II. The
semi-classical quantization of the {\it exact} (circular) string solutions
and the canonical quantization in the string perturbation series
approach of the generic strings,  give the same results.
\vskip 48pt
\hspace*{-6mm}{\bf Acknowledgements:}\\
A.L. Larsen is supported by the Danish Natural Science Research
Council under grant No. 11-1231-1SE

\newpage
\begin{centerline}
{\bf Figure Captions}
\end{centerline}
\vskip 24pt
\hspace*{-6mm}Fig.1. The potential $V(r)=r^2 a(r)$ introduced
after Eq.(2.3) for a circular
string in the three spacetimes: (a) Minkowski spacetime, (b) de Sitter
spacetime, (c) anti de Sitter spacetime.
\vskip 12pt
\hspace*{-6mm}Fig.2. The energy and pressure, Eqs.(2.14), (2.15), for a string
expanding from $r=0$ towards infinity (unstable string)
in de Sitter spacetime. The
curves are drawn for the case $H^2b\alpha'=0.3.$
\vskip 12pt
\hspace*{-6mm}Fig.3. The energy and pressure, Eqs.(2.14), (2.15), for an
oscillating (stable)
string in de Sitter spacetime. The curves describe one period of
oscillation in the case $H^2b\alpha'=0.15.$
\vskip 12pt
\hspace*{-6mm}Fig.4. Parametric plot of $H^2\alpha'W$ as a function of
$H^2m^2\alpha'^2,$ Eqs.(3.20), (3.23), for $k\in[0,\;1]$ in de
Sitter spacetime. Notice that
$H^2m^2\alpha'^2\in
[0,\;1]$ and $H^2\alpha'W\in[0,\;2\sqrt{2}-2\log(1+\sqrt{2})].$ For
$W=2\pi n$ ($n\geq 0$)
there can only be a {\it finite} number of states.
\vskip 12pt
\hspace*{-6mm}Fig.5. Parametric plot of $H^2\alpha'W$ as a function of
$H^2m^2\alpha'^2,$ Eqs.(3.44), (3.46), in anti
de Sitter spacetime. Notice that
$H^2m^2\alpha'^2\in
[0,\;\infty[\;$ and $H^2\alpha'W\in[0,\;\infty[\;.$ For
$W=2\pi n$ ($n\geq 0$)
there are {\it infinitely many} states.
\newpage
\begin{centerline}
{\bf Table Captions}
\end{centerline}
\vskip 24pt
\hspace*{-6mm}Table I. Circular string
energy and pressure in Minkowski, de Sitter and anti de
Sitter spacetimes.
\vskip 24pt
\hspace*{-6mm}Table II. Semi-classical quantization of oscillating
circular strings in Minkowski, de Sitter and anti de
Sitter spacetimes.
\end{document}